\documentclass[a4paper,aps,prb,superscriptaddress,twocolumn]{revtex4}
\usepackage[latin1]{inputenc}
\usepackage{amsmath}
\usepackage{txfonts}
\usepackage{bm}
\usepackage{graphicx}
\newcommand{\ie}{i.~e.,} 

\newcommand{\hc}{{\text{H.\ c.}}}  
\newcommand{\ket}[1]{|#1\rangle}

\newcommand{\matel}[3]{\langle#1|\,#2\,|#3\rangle} 

\newcommand{\gc}{{\mathcal G}_2} 
\newcommand{\gset}{G_{}^S}
\newcommand{\gsym}{{G_0^S}}
\newcommand{\cd}{c_{d}^{\phantom{\dagger}}} 
\newcommand{\fzero}{f_{0}^{\phantom{\dagger}}} 
\newcommand{\fzerod}{f_{0}^{\dagger}} 
\newcommand{\cdd}{c_{d}^\dagger}

\newcommand{\ed}{\varepsilon_d}
\newcommand{\ea}{\varepsilon_a}

\newcommand{\lp}{\bm{\left(}} 
\newcommand{\rp}{\bm{\right)}}
\newcommand{\ha}{H_A}

\newcommand{\mc}[1]{\mathcal{#1}}
\newcommand{\vw}{W}

\newcommand{\za}{\mathcal{Z}}

\newcommand{\gammaw}{\Gamma_{w}}
\newcommand{\deltaw}{\delta_w}

\newcommand{\te}[1]{$\times10^{-#1}$}

\begin{document}
\title{Universal zero-bias conductance through a quantum wire side-coupled
  to a quantum dot}
\author{A. C. Seridonio}
\altaffiliation{Present address:
  Instituto de Física\\ Universidade Federal Fluminense, Niter\'oi,
  24210-346, RJ- Brazil}
\affiliation{Departamento de F\'{\i}sica e
  Inform\'{a}tica, Instituto
  de F\'{\i}sica de S\~{a}o Carlos, \\ Universidade de S\~{a}o Paulo,
  Cx.~Postal~369, 13560 S\~{a}o Carlos, SP, Brazil}

\author{M. Yoshida} \affiliation{Departamento de F\'{\i}sica,
  Instituto de Geoci\^{e}ncias
  e Ci\^encias Exatas,\\ Universidade Estadual Paulista, 13500 Rio
  Claro, SP, Brazil}
\author{L.~N.~Oliveira}
\affiliation{Departamento de F\'{\i}sica e
  Inform\'{a}tica, Instituto
  de F\'{\i}sica de S\~{a}o Carlos, \\ Universidade de S\~{a}o Paulo,
  Cx.~Postal~369, 13560 S\~{a}o Carlos, SP, Brazil}
\begin{abstract}
  A numerical renormalization-group study of the conductance through a
  quantum wire side-coupled to a quantum dot is reported. The
  temperature and the dot-energy dependence of the conductance are
  examined in the light of a recently derived linear mapping between
  the Kondo-regime temperature-dependent conductance and the universal
  function describing the conductance for the symmetric Anderson model
  of a quantum wire with an embedded quantum dot. Two conduction
  paths, one traversing the wire, the other a bypass through the
  quantum dot, are identified. A gate potential applied to the quantum
  wire is shown to control the flow through the bypass. When the
  potential favors transport through the wire, the conductance in the
  Kondo regime rises from nearly zero at low temperatures to nearly
  ballistic at high temperatures. When it favors the dot, the pattern
  is reversed: the conductance decays from nearly ballistic to nearly
  zero. When the fluxes through the two paths are comparable, the
  conductance is nearly temperature-independent in the Kondo regime,
  and a Fano antiresonance in the fixed-temperature plot of the
  conductance as a function of the dot energy signals
  interference. Throughout the Kondo regime and, at low temperatures,
  even in the mixed-valence regime, the numerical data are in
  excellent agreement with the universal mapping.
\end{abstract}

\maketitle

\section{Introduction}
\label{sec:intro}
The transport properties of nanostructured devices constitute a
complex subject. Amongst its numerous facets, interference is one of the
shiniest, because it challenges classical intuition, adds colorful features to
experimental plots, and can be controlled. Such characteristics
have motivated numerous studies, in either the Aharonov-Bohm or the
$T$-shaped interferometric
arrangement.\cite{HKS01:156803,AKS+04:176802,AEO+06:195329,%
  FBI+06:205326,SSI+06_096603,SIS08:153304,BP01:5128,CWB01:4636,%
  THC+02:085302,FFA03:155301,MNU04:3239,KAS+04:035319,%
  SAK+05:066801,ZB06.035332,MSU06:938,Kat07:233201} The minimal number
of components in and the rich variety of effects stemming from the
latter setup make it particularly interesting.

The $T$-shaped device comprises a quantum-dot side-coupled to a
quantum wire, through which current flows in response to a small
bias. Its dynamics parallels that of the single-electron transistor,
the alternative arrangement that embeds the quantum dot in the
conduction path. The same Hamiltonian models the two devices, the
thermodynamical properties are hence identical. Nonetheless, the
conductances in the two geometries are remarkably distinct.

In either geometry, if a gate potential $V_d$ applied to the quantum
dot tunes its occupation $n_d$ to an odd integer, the resultant
magnetic moment interacts antiferromagnetically with the wire
electrons and induces a Kondo cloud. Thin at first, the cloud
gradually shrouds the quantum dot as the temperature is reduced. Below
the Kondo temperature $T_K$, the cloud and dot moment lock into a
singlet. In energy space, the Kondo resonance in the low-energy spectrum of the
model Hamiltonian identifies the consolidation of the Kondo singlet.\cite{hewson93}

The conductances of the single-electron transistor and of the
$T$-shaped devices respond differently to the Kondo resonance. In the
embedded arrangement, the resultant strong coupling between the
conduction and the dot electrons breaks through the Coulomb blockade
to allow conduction. This remarkable low-temperature property of the
single-electron transistor found precise mathematical expression in
the early numerical renormalization-group (NRG) study by Costi,
Hewson, and Zlatic,\cite{CHZ94.19} who described the
temperature-dependent conductance for the spin-degenerate Anderson
model\cite{An61:41} and showed that the conductance for the
particle-hole symmetric model is a monotonically decreasing universal
function $\gset(T/T_K)$ of the temperature scaled by the Kondo
temperature.

In the side-coupled geometry, instead of enhancing the conductance,
the Kondo droplet obstructs transport through the wire. While the thin
high-temperature cloud offers little resistance to electronic flow,
low temperatures raise a Kondo blockade in the wire sector closest to
the dot. In experimental arrangements allowing conduction only through
the wire, the conductance through the $T$-shaped device therefore
rises with temperature. For the symmetric Anderson Hamiltonian, in
particular, the conductance $G$ is complementary to the universal function
$\gset$: $G(T/T_K)=\gc-\gset(T/T_K)$, where $\gc\equiv 2e^2/h$ is the
quantum conductance.\cite{MSU06:938,SYO2009} To emphasize the
significance of this theoretical finding, we define the complementary
symmetric conductance $\gsym(T/T_K)=\gc-\gset(T_K)$.

More ellaborate $T$-shaped setups accommodate an alternative
conduction path through the quantum dot, a bypass around the blocked
wire segment.\cite{KAS+04:035319,SAK+05:066801,KSA+06:36} Considered
on its own, the bypass is equivalent to a single-electron transitor:
while a Coulomb blockade bars electronic transfer through the dot at
high temperatures, the Kondo resonance allows conduction at low $T$.

Most frequently, the two conduction paths coexist. Three conduction
patterns then stem from the combination of the currents through
them.\cite{KAS+04:035319,SAK+05:066801,KSA+06:36,OAK+07:084706} If the
experimental conditions favor the bypass (the path traversing the
wire), the conductance will decay (rise) with temperature,
following the pattern set by the universal (the
complementary-conductance) function $\gset(T/T_K)$ [$\gsym(T/T_K)$].

The third pattern arises when the amplitudes for electronic transfer
along the two paths are comparable. The conductance is then a
relatively flat function of temperature, but the low-temperature
conductance profile (\ie\ $G$ as a function of the dot gate
potential $V_d$ at fixed temperature) shows Fano antiresonances
indicative of interference between the two currents. Under a variable
gate potential applied to the quantum wire, the response of a
$T$-shaped device to a small bias can be adjusted to bring out any of
the three patterns.\cite{KSA+06:36,OAK+07:084706}

Together with such experimental findings, theoretical analyses have
contributed to our understanding of the $T$-shaped device. Important
results have been derived.\cite{HKS01:156803,%
  MNU04:3239,THC+02:085302,FFA03:155301,MSU06:938,SIS08:153304} More
recently, the present authors showed that the temperature-dependent
Kondo-regime conductance $G(T/T_K)$ can be mapped onto the universal
function $\gset(T/T_K)$.\cite{SYO09:000} The mapping is linear, its
coefficient an universal function of the ground-state phase shift
$\delta$ of the wire electrons. The conductance depends on the model
parameters only by way of $\delta$ and $T_K$. If the model parameters
make $\delta=0$ ($\delta=\pi/2$), in particular, it results that
$G(T/T_K)=\gset(T/T_K$) [$G(T/T_K)=\gsym(T/T_K)$].\cite{SYO09:000}

Here, to bring to light the physical content of the universal mapping,
we present essentially exact numerical-renormalization group (NRG)
results for the conductance of the $T$-shaped device as a function of
the dot gate potential and temperature for various wire gate
potentials. In correspondence with the experimental data, the
numerical data display three patterns: the temperature dependence of
the conductances can be close to one of the two universal functions
$\gset(T/T_K)$ and $\gsym(T/T_K)$, or approximately constant; the
corresponding low-temperature conductance profiles are uniformly close
to zero, uniformly close to ballistic, or display Fano antiresonances,
respectively.

The mapping to the universal function $\gset(T/T_K)$ describes
accurately each calculated temperature-dependent conductance in the
Kondo regime and thus condenses in a simple expression the variety of
thermal and dot-energy dependences in the numerical study. It shows,
moreover, that much like the optical phase difference controls the
fringes in a Michelson-Morley interferometer, the ground-state phase
shift controls the interference between the bypass and the cross-wire
currents. For $\delta\approx0$ ($\delta\approx\pi/2$), the former
(latter) is dominant, and $G(T/T_K)\approx\gset(T/T_K)$
[$G(T/T_K)\approx\gsym(T/T_K)$]. For intermediate phase shifts, the
interference between the two currents raises antiresonances. Since
$\delta$ depends on the gate potentical applied to the wire, the three
conduction patterns appear in succession as the potential grows.

This paper closes a quartet dedicated to the conductance for the
Anderson model of a wire coupled to a quantum
dot. References~\onlinecite{SYO2009}~and \onlinecite{SYO2009ii}
discussed the embedded geometry, the former paper having derived the
mapping between the conductance and the universal function
$\gset(T/T_K)$, and the latter having presented NRG data to illustrate and to
probe the accuracy of that mapping at the limits of the Kondo regime
and beyond them. Reference~\onlinecite{SYO09:000} derived the mapping
for the $T$-shaped device, and this report compares it with an NRG
survey of the side-coupled geometry.

Our presentation is divided in 5 sections. Section~\ref{sec:model}
defines the model Hamiltonian and recalls basic concepts associated
with it. The mapping is discussed in Section~\ref{sec:univ}. The
comprehensive overview of the numerical results and comparison with
the mapping to the universal function follow, in
Section~\ref{sec:num}. Finally, Section~\ref{sec:concl} summarizes our
conclusions.
\section{Model}
\label{sec:model}
The overview of the numerical results in Sec.~\ref{sec:num} will
divide the parametrical space of the model in a number of regimes,
easily identified by their nearly uniform conductances. Before
examining the NRG results, it is therefore convenient to discuss the
characteristic energies marking the boundaries of those regimes. This
section defines the model Hamiltonian, relates the conductance to its
eigenvalues and eigenvectors, and refers to the pioneer investigations
that identified the characteristic scales in its
spectrum.\cite{Hal78:416,KWW80:1003,KWW80:1044}
\subsection{Hamiltonian}
\label{sec:mod-ham}
\begin{figure}[th]
  \centering
  \includegraphics[width=0.95\columnwidth]{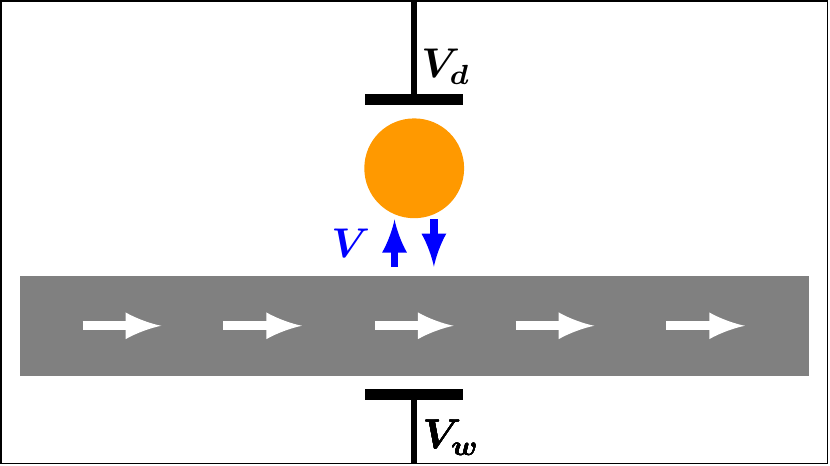}
  \caption[$T$-shape]{(Color online) $T$-shaped device. The circle depicts a quantum dot
    side-coupled to a quantum wire. The gate potentials $V_d$ and
    $V_w$ control the electronic occupations of the dot and of the
    Wannier orbital~(\ref{eq:fzero}), respectively. The device
    allows interference between the current flowing along the primary
    conduction path, indicated by the white arrows, and dot bypass,
    which circumvents the segment of the wire closest to the dot.}
  \label{fig:1}
\end{figure}

The quantum dot in Fig.~\ref{fig:1} is side-coupled to a quantum
wire. Current flows in response to a bias voltage applied to the
wire. Gate potentials $V_d$ and $V_w$ control the dot occupation and
the occupation of the Wannier orbital
\begin{equation}
  \label{eq:fzero}
  \fzero\equiv\frac{1}{\sqrt N}\sum_k a_k,
\end{equation}
where $N$ is the number of conduction states in the wire.

In standard notation, the Anderson Hamiltonian capturing the physics
of this setup is
\begin{equation}
  \label{eq:hand}
  \ha=\sum_{k}\epsilon_k a_{k}^\dagger a_{k} +\vw \fzerod \fzero
  +V(\fzerod \cd+\hc)+H_{d},
\end{equation}
where the states $a_{k}$ form a structureless half-filled conduction
band of halfwidth $D$ and density of states $\rho=N/2D$, the
scattering potential $\vw$ is controlled by the gate voltage $V_w$,
and the tunneling amplitude $V$ couples the spin-degenerate dot level
$\cd$ to the $\fzero$ orbital. 

It proves convenient to write the dot Hamiltonian $H_d$ in
Eq.~(\ref{eq:hand}) in the form
\begin{equation}
  \label{eq:hdot}
  H_d=(\ed+\frac{U}2) n_d - \frac U2 n_{d\uparrow}n_{d\downarrow},
\end{equation}
which defines the antisymmetric component
$\ea\equiv\ed+U/2$ of the dot energy, \ie\ the component
that changes sign under the particle-hole transformation
$\cd\to-\cdd$, $a_k\to a_k^\dagger$. 
The second term on the right-hand side of the Eq.~(\ref{eq:hdot}), by
contrast, remains invariant under the same transformation. For
$\vw=\ea=0$, only the second term survives, and the Anderson
Hamiltonian reduces to its {\em symmetric} form.\cite{KWW80:1003}

To highlight interference, many a study of transport through
a nanodevice has explicitly introduced two conduction paths in the
model Hamiltonian. The ratio between the amplitudes for charge
transfer along the two routes, \ie\ the Fano factor $q$, is then a
model parameter. 

An alternative approach is possible. A sufficiently comprehensive
model Hamiltonian may implicitly include two conduction routes and
hence define an effective Fano factor $q_{eff}$.\footnote{Even if the
  Fano factor is defined with the model Hamiltonian, electronic
  correlations may renormalize it. An example was provided by
  L.~N.~Oliveira and J.~W.~Wilkins, Phys.~Rev.~B, {\bf32}, 696
  (1985).} Fano's pioneer analysis of self-ionization in the He atom
provided the first example of this approach,\cite{Fa61:1866} which we adopt.
A simple extension of Eq.~\eqref{eq:hand} would provide another
example.\cite{SYO09:000} A momentum-dependent coupling $\sum_k
(V_ka_k^\dagger\cd+\hc)$ between the quantum wire and the quantum dot
would introduce a conduction path alternative to transport straight through
the quantum wire and define an effective Fano factor proportional to
the Fermi-level value of the derivative $dV_k/dk$.\cite{SYO09:000} 

The momentum-independent coupling in Eq.~(\ref{eq:hand}) rules out
this alternative route. Section~\ref{sec:num} will show, however, that
the wire gate potential $\vw$ opens a bypass through the quantum dot,
so that an effective Fano factor $q_{eff}$ arises, which grows with
$\vw$. As long as $\vw=0$, Eq.~\eqref{eq:hand} comprises two
conduction paths, through which currents can flow and interference.

To reproduce the model Hamiltonian in Ref.~\onlinecite{SYO09:000}, a
momentum-dependent coupling could have been substituted for the third
term on the right-hand side of Eq.~(\ref{eq:hand}). Any of a variety
of artificial couplings between the dot and the wire orbitals
orthogonal to $f_0$ might equally well have been introduced, to
generate another bypass and originate a different Fano parameter. In
any case, the results would be equivalent to those in
Sec.~\ref{sec:num}, because the universal mapping between the
temperature-dependent conductance and the function $\gset(T/T_K)$
derived in Ref.~\onlinecite{SYO09:000} depends only on the
ground-state phase shift $\delta$. Since $\delta$ decreases from
$\pi/2$ to zero as the wire potential $\vw$ grows, our survey of the
conductance for the structure in Fig.~\ref{fig:1} samples well the
dependence of the conductance on Fano factors. Alternative
Hamiltonians would yield the same information.

\subsection{Conductance}
\label{sec:mod-cond}

Reference~\onlinecite{SYO2009} derived an expression for the
temperature-dependent electrical conductance $G(T)$ in the embedded
configuration.  The same analysis leads to an analogous equality for
the conductance in the side-coupled configuration:
\begin{equation}
  \label{eq:thermalG}
      G(T) = \gc\,\frac{\beta\pi\Gamma}{\za}\,
\sum_{mn}\frac{|\matel{m}{\fzero}{n}|^2}{e^{\beta E_m}+e^{\beta E_n}},
\end{equation}
where $\za$ is the partition function, $\ket{m}$ ($\ket{n}$) is an
eigenstate of $\ha$, with eigenvalue $E_m$ ($E_n$), and $\Gamma\equiv \pi\rho
V^2$ is the dot-level width, due to its coupling to the quantum wire.

To describe the temperature and dot-energy dependence of the
conductance, it is sufficient to consider non-negative wire gate
potentials. Since the model Hamiltonian and the conductance remain
invariant under the modified particle-hole transformation
\begin{equation}
  \label{eq:phtransf}
  \begin{array}[c]{rr}
a_k\to& a_k^\dagger,\\
\cd\to&-\cdd,\\
\ea\to&-\ea,\\
\text{and}\\
\vw\to&-\vw,
  \end{array}
\end{equation}
given two energies $\alpha <0$ and $\beta$, the conductance for
$\vw=\alpha$ and $\ea=\beta$ is given by the right-hand side of
Eq.~(\ref{eq:thermalG}) for $\vw=-\alpha>0$ and $\ea=-\beta$. The
conductance for a negative wire gate potential can therefore be
obtained from the results for the symmetric (positive) potential.

\subsection{Characteristic energy scales}
\label{sec:mod-char}
Figure~3 in Ref.~\onlinecite{SYO2009} depicts the five characteristic
energies in the spectrum of the Hamiltonian~(\ref{eq:hand}). The first
(second), due to the dot Hamiltonian~(\ref{eq:hdot}), is the
excitation energy $\Delta_0=-\ed$ ($\Delta_2=U+\ed$) necessary to
remove an electron from (add an electron to) the singly-occupied
level. The only scale in the conduction band Hamiltonian is trivial:
the halfwidth $D$. The coupling to the dot, on the other hand, makes
two important contributions: the level width $\Gamma$, and the Kondo
energy $k_BT_K$. The scattering potential $\vw$ reduces the former to
\cite{SYO2009}
\begin{equation}
  \label{eq:gammaw}
  \gammaw = \frac{\pi\rho V^2}{1+\lp\pi\rho \vw\rp^2}.
\end{equation}

The Kondo regime is defined by the condition $\max(k_BT, \gammaw)\ll
\mc{E}_{c}^K$, where the dominant characteristic energy is
$\mc{E}_c^K= \min(D, \Delta_0^*, \Delta_2^*)$. The $n_d=0$ and $n_d=2$
dot configurations are then energetically inaccessible, and a dot
magnetic moment arises. As the temperature is reduced below the
band halfwidth $D$, the antiferromagnetic interaction between the
conduction electrons and the dot electron progressively screens the
moment. The energy of the resulting low-temperature singlet defines
the fifth characteristic scale in our set: the Kondo energy $k_BT_K$.

A side effect of the coupling between the dot and the conduction
electrons, one that emerges in the spectrum of particle-hole
asymmetric Hamiltonians, is the renormalization $\Delta_0\to
\Delta_0^*$, or $\Delta_2\to\Delta_2^*$, of the lowest dot-excitation
energy.\cite{Hal78:416,KWW80:1044} Although of no other practical
consequence, this energy shift is important because it displaces both
the Kondo and the mixed-valence regimes.\cite{SYO2009ii}

The latter regime comprises the two regions in the parametric space of the
model Hamiltonian that enclose the Kondo regime. One of them is
defined by the inequality $\gammaw\agt \Delta_0^*$, the other, by
$\gammaw\agt \Delta_2^*$. In the mixed-valence regime the dominant
characteristic energy is $\mc{E}_c^{\text{m-v}}=\gammaw$, and at high
temperatures, the dot moment is only partially formed. As the thermal
energy is reduced past $\gammaw$, the dot level $\cd$ couples strongly
to the surrounding conduction electrons, and all physical properties
approach their low-temperature limit.

\section{Universal mapping}
\label{sec:univ}

The derivation of the universal expression mapping the conductance in
Eq.~(\ref{eq:thermalG}) to the universal function $\gset(T/T_K)$
computed by Costi, Hewson, and Zlatic\cite{CHZ94.19,BCP08:395} has
been summarized elsewhere.\cite{SYO09:000} A detailed presentation of
the analogous derivation for the single-electron transistor being moreover
available,\cite{SYO2009} only recapitulation of the central result is
required:
\begin{equation}
  \label{eq:guniv}
    G\lp\frac{T}{T_K}\rp -\frac{\gc}2 =
    \lp\gset\lp\frac{T}{T_K}\rp-\frac{\gc}2\rp\cos(2\delta).
\end{equation}
Here, $\delta$ is the ground-state phase shift of the conduction
electrons in the wire.

To make the physical content of Eq.~(\ref{eq:guniv}) more visible, we
rewrite it in the equivalent form
\begin{equation}
  \label{eq:univInterf}
  G\lp\frac{T}{T_K}\rp = \gset\lp\frac{T}{T_K}\rp\,\cos^2\delta
+ \gsym\lp\frac{T}{T_K}\rp\sin^2\delta,
\end{equation}
which shows that the conductance is a linear combination of the
universal function $\gset(T/T_K)$ with its complement,
$\gsym(T/T_K)$. While the universal, monotonically decreasing function
$\gset(T/T_K)$ describes the conductance through a particle-hole
symmetric quantum dot in the embedded configuration, $\gsym(T/T_K)$ is
the monotonically increasing function that describes the conductance
through a particle-hole symmetric quantum dot in the side-coupled
configuration. For particle-hole asymmetric $T$-shaped devices, the
first term on the right-hand side of Eq.~(\ref{eq:univInterf})
describes the current through the bypass, while the second term
accounts for the current traversing the wire.  If the two conduction
paths were independent, the resultant conductance would be the sum
$G_d(T/T_K)+G_w(T/T_K)=\gc$; the coefficients $\cos^2\delta$ and
$\sin^2\delta$ thus account for the interference between the two fluxes.

\subsection*{Special limits}
\label{uni-special}

\begin{subequations}
  \label{eq:g4}
  Four limits of Eq.~\eqref{eq:univInterf} merit
  special attention. For $\delta=\pi/2$, the mapping reduces to
  \begin{equation}
    \label{eq:gzero}
    G(\frac{T}{T_K}) = \gc -
    \gset(\frac{T}{T_K})\equiv\gsym(\frac{T}{T_K})
    \qquad(\delta=\pi/2).
  \end{equation}

  The equality, which can be derived from a diagrammatic expansion for
  the symmetric Anderson Hamiltonian,\cite{MNU04:3239,SYO2009} shows
  that, with $\delta=\pi/2$, the conductances for the side-coupled and
  the embedded geometries are complementary. This result is
  particularly important because, in the absence of a wire gate
  potential, the Friedel sum rule\cite{La66:516} pins the
  ground-state phase shift to the vicinity of $\pi/2$.\cite{SYO2009ii} For $\vw=0$,
  the conductance through the $T$-shaped device is therefore expected
  to decay from nearly ballistic to nearly zero as the temperature
  rises.

  In the second special limit, $\delta=0$, Eq.~(\ref{eq:guniv}) yields
  \begin{equation}
    \label{eq:gset}
    G(\frac{T}{T_K})=\gset(\frac{T}{T_K})\qquad(\delta=0).
  \end{equation}
  In the absence of a wire gate potential, the ground-state phase
  shift is close to $\pi/2$; a strong wire potential $\vw$ must be
  applied to make $\delta\approx0$. The potential affects the
  conductance in two different ways: it induces a charge that blocks
  transport through the central region of the wire; at the same time,
  it opens a bypass through the quantum dot. At high temperatures, the
  Coulomb blockade against conduction through the dot makes the
  bypass ineffective, so that the conductance of the device
  vanishes. Upon cooling, however, the Kondo effect raises the
  blockade and the conductance rises, lifted by the very same
  mechanism that allows conduction through a single-electron
  transistor.\cite{GR87:452,NL88.1768,GSM+98.156}
  In brief, Eq.~(\ref{eq:gset}) tells us that a strong wire potential
  diverts the current to the bypass and makes the $T$-shaped device
  emulate a single-electron transistor.

  The third simple limit of Eq.~(\ref{eq:guniv}) is $G(T=0)$. At low
  temperatures, the universal curve $\gset$ approaches $\gc$. From
  Eq.~(\ref{eq:guniv}), it follows that
  \begin{equation}
    \label{eq:gfl}
    G(T)= \gc\cos^2\delta\qquad(T\ll T_K),
  \end{equation}
  a result that can be derived from an extension of Langreth's
  argument\cite{La66:516} relating the low-energy dot-level spectral
  density to the ground-state phase shift.\cite{SYO09:000}

  Finally, at high temperatures, the universal function $\gset$
  vanishes, and Eq.~(\ref{eq:guniv}) yields
  \begin{equation}
    \label{eq:glm}
    G(T)= \gc\sin^2\delta\qquad(T\gg T_K),
  \end{equation}
\end{subequations}
Analogous to Eq.~(\ref{eq:gfl}), this result also stems from
Langreth's expression.  In the temperature range $k_BT_K\ll k_B T \ll
\mc{E}^{\text{K}}_c$, within the Kondo regime, the Anderson Hamiltonian
is in the local-moment regime, equivalent to a phase shifted
conduction band weakly interacting with the dot
moment.\cite{KWW80:1003} The interaction neglected, the Hamiltonian
reduces to a free conduction band, to which Langreth's reasoning
applies.\cite{SYO09:000,SYO2009} The phase shift $\delta_{LM}$ of the
wire electrons can be obtained from the Friedel sum rule, which
associates a phase-shift difference $\pi/2$ with the Kondo cloud. We
hence have that $\delta=\delta_{LM}+\pi/2$, and so
$\delta_{LM}=\delta-\pi/2$.  Substitution of $\delta_{LM}$ for
$\delta$ in Eq.~(\ref{eq:gfl}) leads to Eq.~(\ref{eq:glm}).

\section{Numerical results}
\label{sec:num}
We are now ready to discuss the NRG results and compare them with
Eq.~(\ref{eq:guniv}). A previous report detailed the numerical
procedure yielding the ground-state phase shift $\delta$, and the
temperature-dependent conductance $G(T)$ and tabulated the parameters
controlling the accuracy of the computation.\cite{SYO2009ii}
Here, to follow the structure of that paper, we will discuss $\delta$
before examining $G(T)$.

\subsection{Phase shifts}
\label{sec:num-phas}

\begin{figure}[th]
  \centering
  \includegraphics[width=0.95\columnwidth]{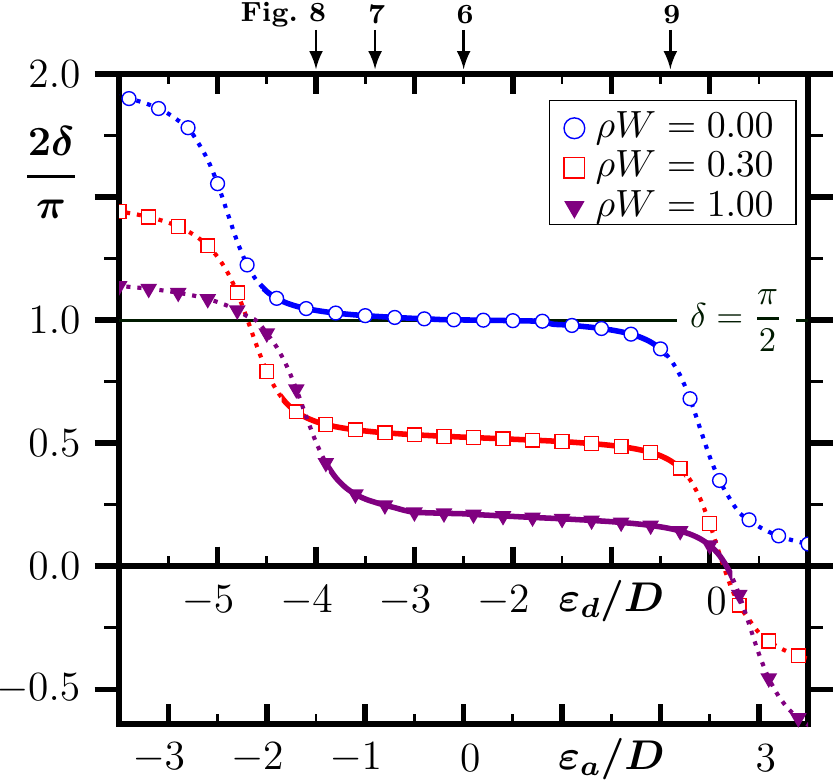}
  \caption{(Color online) Phase shifts as functions of the dot-level
    energy for the indicated wire potentials $W$. The triangles,
    squares, and circles were obtained from the low-energy fixed-point
    Hamiltonian for $U=5\,D$ and $\gammaw=0.15\,D$. For each curve, a
    solid line guides the eye through the data points in the pertinent
    Kondo regime; the dotted lines guide the eyes elsewhere. The
    horizontal line labeled $\delta=\pi/2$ allows identification of
    the dot-energies at which the right-hand side of
    Eq.~(\ref{eq:gfl}) vanishes. The vertical arrows pointing to the
    top axis indicate the dot energies at which the temperature
    dependent conductance is shown in Figs.~6-9.}
  \label{fig:2}
\end{figure}

Figure~\ref{fig:2} shows the ground-state phase shift, extracted from
the low-energy spectrum of the Hamiltonian~$\ha$, as a function of the
dot-energy $\ed$ in the interval $|\ed+U/2|< 3.5\,D$, \ie\ for
$|\ea|<3.5\,D$.  It follows from the Friedel sum rule that the
ordinate $2\delta/\pi$ is the extra charge $n_w$ piled up at the
wire. That accumulation is the sum of two charges: the charge induced by the
wire potential, the absolute value of which grow with $\vw$; and the
Kondo screening charge, which is equal to the dot occupation $n_d$ and
hence close to unity in the Kondo regime. 

With no wire potential, the charges $n_w$ and $n_d$ coincide.  If
$\ea=0$, then $\ha$ reduces to the symmetric Hamiltonian, the dot
occupation is unitary, and $\delta=\pi/2$. If, on the other hand,
$\ea\ne0$, then a particle-hole transformation, which reverses the
sign of $\ea$, transmutes $n_d$ into $2-n_d$ and hence changes the
sign of $\delta-\pi/2$. The curve through the circles in
Fig.~\ref{fig:2} thus remains invariant under a rotation of 180$^0$
around the point $\ea=0$, $\delta=\pi/2$.

The solid line through the circles marks the Kondo regime, throughout
which $\delta\approx\pi/2$. Comparison with the solid lines through
the squares and triangles shows that the positive gate potential
applied to the wires displaces the Kondo regime to higher dot
energies.

The negative charge induced by the potential pushes down the phase
shifts and spoils the symmetry about $\ea=0$, $\delta=\pi/2$. In the
region $\ed>0$, the wire charge $n_w$ can now become negative. While
the circles cross the $\delta=\pi/2$ line at a saddle point, the slope
of the curves through the squares and triangles is markedly
negative. Thus, in contrast with the circles, only in narrow ranges of
dot energies do the squares and triangles dwell near $\delta=\pi/2$ or
$\delta=0$. As the data moreover show, except for $\vw\approx0$, the
range in which $\delta\approx0$ ($\delta\approx\pi/2)$ is pinned to
the vicinity of the mixed-valence region $\ea\approx U/2$
($\ea\approx-U/2$).

Combined with Eq.~(\ref{eq:guniv}), the information in
Fig.~\ref{fig:2} determines the thermal dependence of the conductance
through the device in Fig.~\ref{fig:1}. Detailed comparison between
the resulting curves and the NRG conductances will be presented in
Secs.~\ref{sec:num-phas}-\ref{sec:num-strong}. Before that, to provide
a more visual description of the numerical data, we present bird's-eye
views of the conductance as a function of the temperature and
dot-level energy.

\subsection{Conductance for $\vw=0$}
\label{sec:num-zero}
\begin{figure}[th]
  \centering
  \includegraphics[width=0.90\columnwidth]{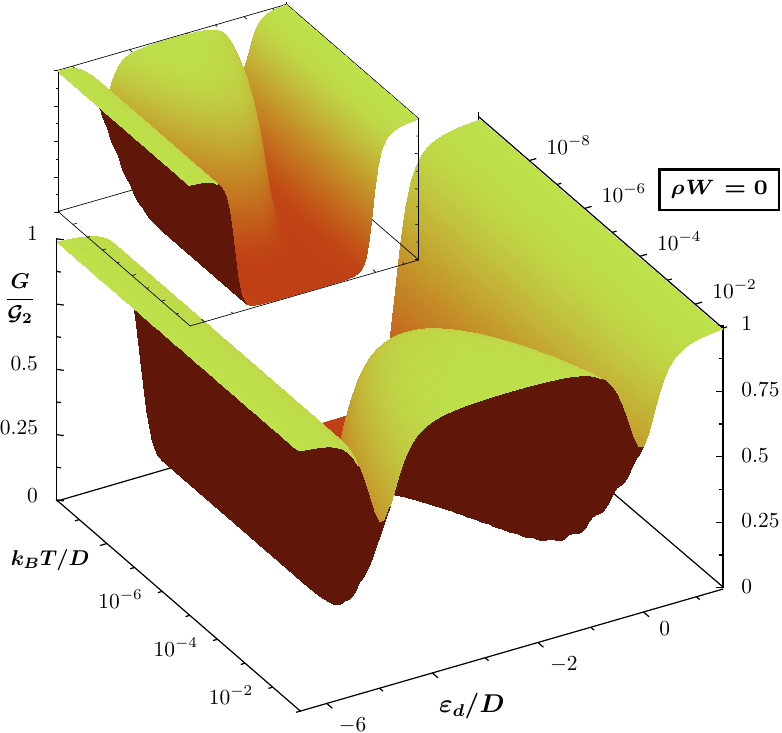}
  \caption{(Color online) Conductance as a function of the temperature and dot-level
    energy for $\vw=0$, $U=5\,D$, and $\gammaw=0.15\,D$. The inset
    shows the same plot from the opposite viewpoint, \ie\ so that,
    instead of rising, the temperature decays toward the
    viewer. The invariance of $G(T)$ under particle-hole
    transformations makes the plot symmetric with respect to the
    $\ed=-U/2$ plane. At high temperatures in the Kondo regime, the
    main plot offers an unosbtructed view of the  high-temperature
    plateau, while the Kondo valley at low temperatures can only be
    seen in the inset. The two ridges adjacent to the Kondo regime,
    distinctively marked by the bell-shaped resonances at the frontal
    plane of the main plot, belong to the mixed-valence regime. The
    ballistic flaps at the $\ed=-6\,D$ and $\ed=D$ ends of the
    landscape correspond to the dot occupations $n_d=2$ and $n_d=0$, respectively. }
  \label{fig:3}
\end{figure}

Figure~\ref{fig:3} is a landscape consolidating 71 $G(T)$ curves
calculated for $\vw=0$, $U=5\,D$, $\gammaw=0.15\,D$, and dot energies
in the range displayed in Fig.~\ref{fig:2}. The inset is a
reversed-perspective view of the same plot offering an unobstructed
depiction of the low-temperature region. 

Eqs.~(\ref{eq:gzero})-\eqref{eq:glm} explain the salient features of the
landscape. The central portion of the plot encompasses the Kondo
regime. Here, as the circles in Fig.~\ref{fig:2} show, the
ground-state phase shift is close to $\pi/2$. At high temperatures,
Eq.~(\ref{eq:glm}) yields the nearly ballistic conductance depicted by
the hood-like central plateau visible in the main plot and inset. At
low temperatures, Eq.~(\ref{eq:gfl}) brings the conductance down to
nearly zero; the resulting Kondo valley can be seen in the inset. 

We defer to Sec.~\ref{sec:num-thermal-sym} the quantitative discussion
of the crossover between the plateau and the valley. Here, we note
that the Kondo temperature rises with $|\ea|$. For $|\ea|\approx U/2$,
in particular, \ie\ for $\Delta_0\equiv-\ed\alt\Gamma$ or
$\Delta_2\equiv\ed+U\alt\Gamma$, the Kondo temperature becomes
comparable with $\gammaw$, an indication that the Hamiltonian has
transposed the limits of the Kondo domain to enter the mixed-valence
regime.

At high temperatures, the mixed-valence regime is identified by the
two antiresonances bringing the conductance to $\gc/2$ on the frontal
plane in the main plot, or the rear plane in the inset. As the
temperature is reduced, the conductance descends into the Kondo
valley. As Sec.~\ref{sec:num-zero} will show, the decay can still be
mapped onto the universal function $\gset(T/T_K)$. Given the high
crossover temperature, however, only the low-temperature tail of the
numerical results agrees quantitatively with (\ref{eq:guniv}).

Beyond the mixed-valence regime, as $|\ea|>$ grows past $U/2$, the dot
occupation approaches an even integer, $n_d=0$ or $n_d=2$, the
coupling to the quantum wire becomes ineffective, and the electrons
flow ballistically accross the wire. 

\subsection{Conductance for weak scattering potentials }
\label{sec:num-weak}

\begin{figure}[th]
  \centering
  \includegraphics[width=0.85\columnwidth]{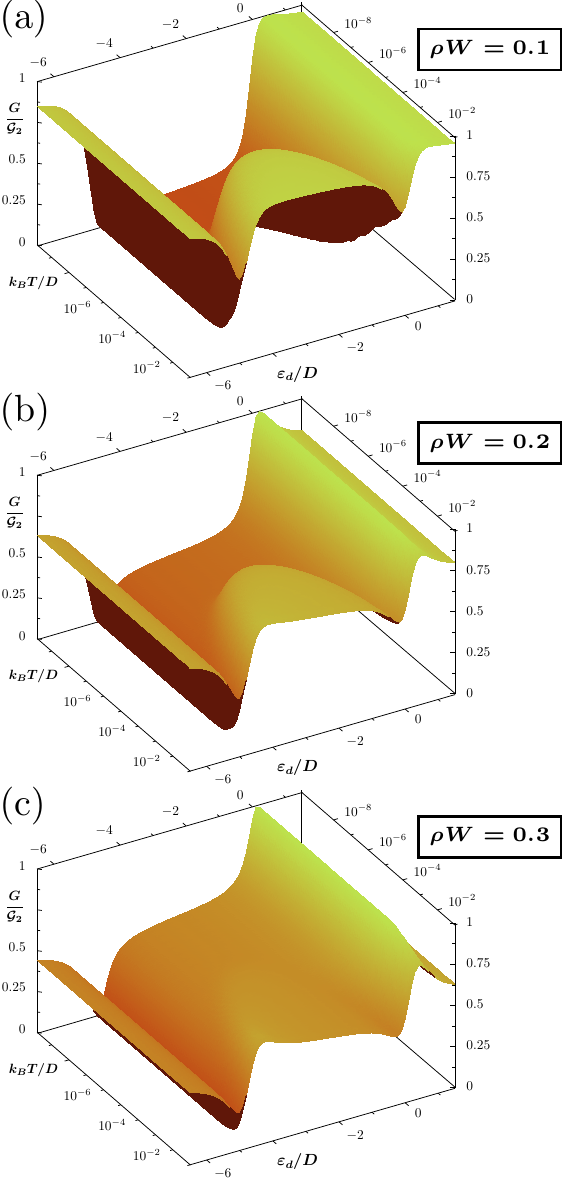}
  \caption{(Color online) Conductance as a function of the temperature
    and dot-level energy for $U=5\,D$, $\gammaw=0.15\,D$, and the
    indicated wire gate potentials $W$. The wire potential reduces the
    conductance at the Coulomb-blockade plateau and raises the
    Kondo-valley conductance. For $\rho W=0.3$ [panel(c)], in the Kondo regime
    the landscape is nearly flat. The wire potential also makes the
    transitions into the mixed-valence regimes markedly asymmetric: at
    fixed temperature below $k_BT=10^{-3}\,D$, the conductance is zero
    at the $\Delta_2^*=0$ and ballistic at the $\Delta_0^*=0$
    resonance. As a result, the low-temperature conductance displays
    the askew profile distinctive of Kondo anti-resonances.}
  \label{fig:4}
\end{figure}

Even moderate gate potentials applied to the wire change qualitatively
the conductance landscape. Figure~\ref{fig:4} shows plots
analogous to Fig.~\ref{fig:3}, calculated for the same Coulomb
repulsion $U=5\,D$ and effective dot-level width $\gammaw=0.15\,D$,
and for three weak potentials: (a) $\rho W=0.1\,D$; (b) $\rho W=0.2\,D$;
and (c) $\rho W=0.3\,D$. Compared to the symmetric landscape in
Fig.~\ref{fig:3}, the new plots show distinctions that evolve rapidly
under the rising gate potential.

The Kondo-valley conductance rises with $\rho W$, while the
high-temperature plateau diminishes. For the highest potential shown,
$\rho W=0.3$, the Kondo-regime conductance is nearly uniform. Outside
the Kondo regime, in the even-$n_d$ regions, the charge $2\deltaw/\pi$
induced by the wire potential lowers the ground-state phase shift to
$-\deltaw$ ($\pi-\deltaw$) in the $n_d=0$ ($n_d=2$) domain. The right-hand
side of Eq.~(\ref{eq:gfl}) is no longer ballistic: $G(T=0)=\gc
\cos^2\deltaw$. 

In the mixed-valence regime, the transition from the
Kondo regime to the $n_d=0$ domain is now markedly different from the
transition to the $n_d=2$ domain. The former encompasses the narrow
$\delta=0$ dot-energy range in Fig.~\ref{fig:2}, for which the
right-hand side of Eq.~(\ref{eq:gfl}) yields $G(T=0)\approx\gc$; while
the latter traverses the equally narrow $\delta=\pi/2$ range, for
which Eq.~(\ref{eq:gfl}) yields $G(T=0)\approx0$. An insulating ravine
develops in the conductance landscape, pinned near $\ed=-U$, while a
ballistic ridge arises near the $\ed=0$ plane. Plotted at fixed
temperature $k_BT \ll\gammaw$ against the dot energy, the conductance
displays the trough-crest pair that defines a Fano
profile.\cite{Fa61:1866} As evidenced by Fig.~\ref{fig:4}(c), this
fiducial mark of interference between currents flowing along parallel
paths becomes most pronounced for intermediate wire potentials.

A brief glance at Fig.~\ref{fig:1} identifies the interfering
fluxes. One of them, indicated by the white arrows, is subject to
obstruction by the charge accumulated in the central segment of the
wire. The second conduction path runs through the quantum dot. To
provide an effective bypass, it must avoid the Wannier orbital
$\fzero$. At first sight, given that Eq.~(\ref{eq:hand}) couples the
quantum dot to $\fzero$, this may seem impossible, and in fact it is,
for $\vw=0$. The wire potential nonetheless spreads the coupling over
wire states beyond the central orbital closest to the dot.

To be more specific, we let $V\to0$ and consider the resulting wire Hamiltonian,
\begin{equation}
  \label{eq:hwire}
  H_w= \sum_k\epsilon_kc_k^\dagger c_k +\frac{W}{N}\sum_{k,k'}c_k^\dagger c_{k'}.
\end{equation}
The diagonalization of this quadratic form yields\cite{SYO2009}
\begin{equation}
  \label{eq:hwdiag}
  H_w = \sum_{\ell}\varepsilon_{\ell}g_\ell^\dagger g_\ell,
\end{equation}
where
\begin{equation}
  \label{eq:gellck}
  g_{\ell}=\sum_{k}\alpha_{\ell,k}c_k,
\end{equation}
with coefficients $\alpha_{\ell, k}$ that depend on $\vw$. The
eigenvalues $\varepsilon_{\ell}$ are phase-shifted with respect to the
$\epsilon_k$. In the vicinity of the Fermi level, in particular, the
phase shifts $\deltaw$ are uniform:\cite{KWW80:1044,SYO2009}
\begin{equation}
  \label{eq:deltaw}
  \tan\deltaw = -\pi\rho \vw.
\end{equation}

In analogy with Eq.~(\ref{eq:fzero}), we can therefore define
\begin{equation}
  \label{eq:phi0}
  \phi_0 \equiv \frac1{\sqrt{N}}\sum_{\ell} g_\ell.
\end{equation}

A straightforward calculation shows that $\{\phi_0^\dagger,
\fzero\}=\cos\deltaw$.\cite{SYO2009} For $\vw=0$, in particular,
$\phi_0=\fzero$. It is also easy to check that the ground-state
occupation $n_0$ of the $\phi_0$ orbital is $1+2\deltaw/\pi$, so that
as required by the Friedel sum rule, the gate potential brings in an
extra charge $2\deltaw/\pi$.

The operator $\phi_0$ plays the role that belonged to $\fzero$ when
$\vw=0$. In the Kondo regime, the antiferromagnetic interaction
between the dot moment and the moment of the electrons occupying the
$\phi_0$ orbital is now responsible for the Kondo
crossover,\cite{KWW80:1044,SYO2009} and at low temperatures, the
$\phi_0$ occupation, equal to the charge in the Kondo cloud plus that
induced by the wire potential, obstructs the transport along the white
arrows in Fig.~\ref{fig:1}. At the same time, the overlap
$\sin\deltaw$ between $\fzero$ and the wire states orthogonal to
$\phi_0$ offers the dot bypass, through which the current runs to
interfere with the flow across $\phi_0$ and generate the askew profile
most visible on the $k_BT=10^{-10}\,D$ plane (\ie\ the rear plane) in
Fig.~\ref{fig:4}(c).

The amplitude for tunneling through the dot is $\sin\deltaw$. The
amplitude for tunneling accross the orbital $\phi_0$ is
$\cos\deltaw$. The ratio between the two amplitudes defines the
effective Fano parameter
\begin{equation}
  \label{eq:qfano}
  q_{eff}\equiv\tan\deltaw=-\pi\rho\vw.
\end{equation}

For $\vw=0$, the right-hand side vanishes. The resulting conductance
curve, stamped on the $k_BT=10^{-10}\,D$ plane in Fig.~\ref{fig:3},
exhibits the symmetry of the $q=0$ Fano profile.\cite{Fa61:1866} For
$\rho\vw=0.3$, by contrast, $q_{eff}\approx-1$, and the conductance
profile, Fig.~\ref{fig:4}(c), is close to the maximum Fano asymmetry.

\subsection{Conductance for strong scattering potentials}
\label{sec:num-strong}

Under stronger gate potentials, while the amplitude for conduction
along the white arrows in Fig.~\ref{fig:1} diminishes, the dot bypass
becomes more effective. The absolute value of the right-hand side of
Eq.~(\ref{eq:qfano}) grows, and the conductance profile at low
temperatures gradually acquires the symmetry of the large-$q$ Fano
profile.

\begin{figure}[th]
  \centering
  \includegraphics[width=0.8\columnwidth]{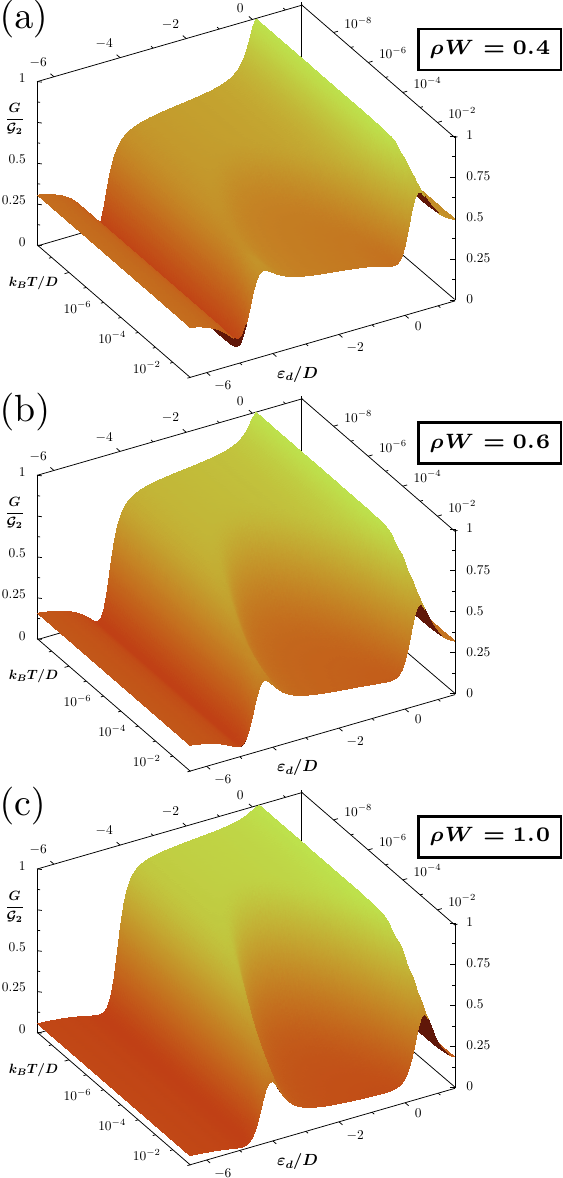}
  \caption{(Color online) Conductance as a function of the temperature
    and dot-level energy for $U=5\,D$, $\gammaw=0.15\,D$, and the
    indicated wire potentials $W$. In the Kondo regime, \ie\ between
    the $\Delta^*_2=0$ and $\Delta_0^*=0$ resonances, as $\rho W$
    grows, the Coulomb-blockade region evolves into a valley, while at
    low temperatures the conductance rises toward a ballistic Kondo
    plateau. Outside the Kondo regime, the conductance approaches
    zero. Overall, the large-$\rho W$ landscape reproduces the
    temperature and dot-level dependence of the conductance through a
    single-electron transistor.}
  \label{fig:5}
\end{figure}

Illustrative landscapes are displayed in Figs.~\ref{fig:5}(a), (b),
and (c), computed for $\rho\vw=0.4$, 0.6, and 1.0, respectively, and
$U=5\,D$ and $\gammaw=0.15\,D$. The dominant features of
Fig.~\ref{fig:4} reappear in the three plots. In particular, the
ballistic ridge and the insulating ravine are still clearly visible in
the mixed-valence regime. Sharp in (a) and (b) the Fano antiresonance
becomes dull in Fig.~\ref{fig:5}(c), because as shown by the triangles
in Fig.~\ref{fig:2}, the wire potential pushes $\delta(\ed)$ so far
down that the phase-shift curve crosses the horizontal lines at the
$\delta=\pi/2$ and $\delta=0$ with relatively small slopes.

In the Kondo regime, while the high-temperature conductance falls, the
low-temperature conductances rises steadily with $\rho W$. In the end,
with $\rho\vw=1$, the conductance is nearly ballistic in the
low-temperature Kondo plateau, while at high temperatures the Coulomb
blockade (the charge induced by the wire potential) impedes transport
through the dot (through the wire) and reduces the conductance to
nearly zero.

The trend in Fig.~\ref{fig:5} indicates that, in the large $\rho\vw$
limit, the conductance landscape is complementary to the plot in
Fig.~\ref{fig:3}, \ie\ $G_{\vw\to\infty}(T)+G_{\vw=0}(T)=\gc$. In view
of Eq.~(\ref{eq:gzero}), we expect $G_{\vw\to\infty}(T)$ to reproduce
the thermal-dependence of the conductance for a single-electron
transistor.\cite{SYO2009,SYO2009ii} Comparison with Fig.~4 in
Ref.~\onlinecite{SYO2009ii}, which describes a single-electron
transistor with the same model parameters, confirms that, with
$\rho\vw\to\infty$, the conductances in the side-coupled and the
embedded configurations are identical. 

For $\vw\to0$, instead of equal, the conductancesa are complementary.
While the conductance for the $T$-shaped device displays the
Kondo-valley profile in Fig.~\ref{fig:3}, the conductance through the
single-electron transistor, which is insensitive to changes in the
wire gate potential, retains the Kondo-plateau pattern to which the
trend in Figs.~\ref{fig:5}(a)-(c) points.

\subsection{Thermal dependence of the conductance}
\label{sec:num-thermal}

This section compares the NRG results for the temperature dependence
of the conductance with the mapping~(\ref{eq:guniv}). We fix the
parameters $U=5\,D$ and $\gammaw=0.15\,D$, and consider the eight gate
wire potentials $\rho\vw$ in Table~\ref{tab:1}. To sample the
dot-energy dependence of the data, four plots will be discussed,
correspondig to the four dot energies indexed by the vertical arrows
in Fig.~\ref{fig:2}. For each run, the table shows the Kondo
temperature resulting from the definition $G(T_K)\equiv\gc/2$ and the
ground-state phase shift calculated from the low-energy eigenvalues of
$\ha$.\cite{SYO2009ii}

\begin{table}[th]
  \centering
  \caption{Phase shifts and Kondo temperatures for the 32 NRG runs
    depicted in Figs.~\ref{fig:6}-\ref{fig:9}. The ground-state phase
    shifts $\delta$ come from the low-energy spectrum of the
    model Hamiltonian, and the Kondo
    temperatures, from the definition $G(T= T_K)\equiv\gc/2$. The
    Kondo temperature marked with asterisks belong to Hamiltonians in
    the mixed-valence regime and, as explained in the text, had to be
    obtained by matching the solid lines to conductances computed at
    temperatures below $T_K$. \label{tab:1}}
  \begin{tabular}{c|crrc|crrc}
    \hline\hline
    Figure & Symbol & $\rho W$ & $\delta/\pi$ & $k_BT_K/D$ &
    Symbol & $\rho W$ & $\delta/\pi$ & $k_BT_K/D$ \\
    \hline
    6&$\circ$&0.00&0.50&8.1\te{7}&$\bullet$&0.10& 0.40&8.6\te{7}\\
    6&$\square$&0.20& 0.32&1.0\te{6}&$\blacksquare$&0.30& 0.26&1.3\te{6}\\
    6&$\lozenge$&0.40& 0.22&1.6\te{6}&$\blacklozenge$&0.60& 0.16&2.5\te{6}\\
    6&$\triangle$&0.80& 0.13&3.8\te{6}&$\blacktriangle$&1.00& 0.11&6.0\te{6}\\
    7&$\circ$&0.00&0.51&4.4\te{6}&$\bullet$&0.10& 0.41&5.9\te{6}\\
    7&$\square$&0.20& 0.33&8.9\te{6}&$\blacksquare$&0.30& 0.27&1.4\te{5}\\
    7&$\lozenge$&0.40& 0.23&2.2\te{5}&$\blacklozenge$&0.60& 0.17&5.7\te{5}\\
    7&$\triangle$&0.80& 0.15&1.4\te{4}&$\blacktriangle$&1.00& 0.13&3.6\te{4}\\
    8&$\circ$&0.00&0.52&8.8\te{5}&$\bullet$&0.10& 0.43&1.4\te{4}\\
    8&$\square$&0.20& 0.35&2.5\te{4}&$\blacksquare$&0.30& 0.29&4.5\te{4}\\
    8&$\lozenge$&0.40& 0.25&9.6\te{4}&$\blacklozenge$&0.60& 0.22&3.0\te{3}\\
    8&$\triangle$&0.80& 0.22&1.0\te{2}${}^*$&$\blacktriangle$&1.00& 0.25&1.2\te{2}${}^*$\\
    9&$\circ$&0.00& 0.42&8.0\te{3}&$\bullet$&0.10& 0.34&4.5\te{3}\\
    9&$\square$&0.20& 0.27&2.7\te{3}&$\blacksquare$&0.30& 0.21&2.0\te{3}\\
    9&$\lozenge$&0.40& 0.17&1.4\te{3}&$\blacklozenge$&0.60& 0.12&6.6\te{4}\\
    9&$\triangle$&0.80& 0.09&3.5\te{4}&$\blacktriangle$&1.00& 0.08&1.9\te{4}\\
    \hline\hline
  \end{tabular}
\end{table}

\subsubsection{Symmetric dot Hamiltonian}
\label{sec:num-thermal-sym}

\begin{figure}[th]
  \centering
  \includegraphics[width=0.95\columnwidth]{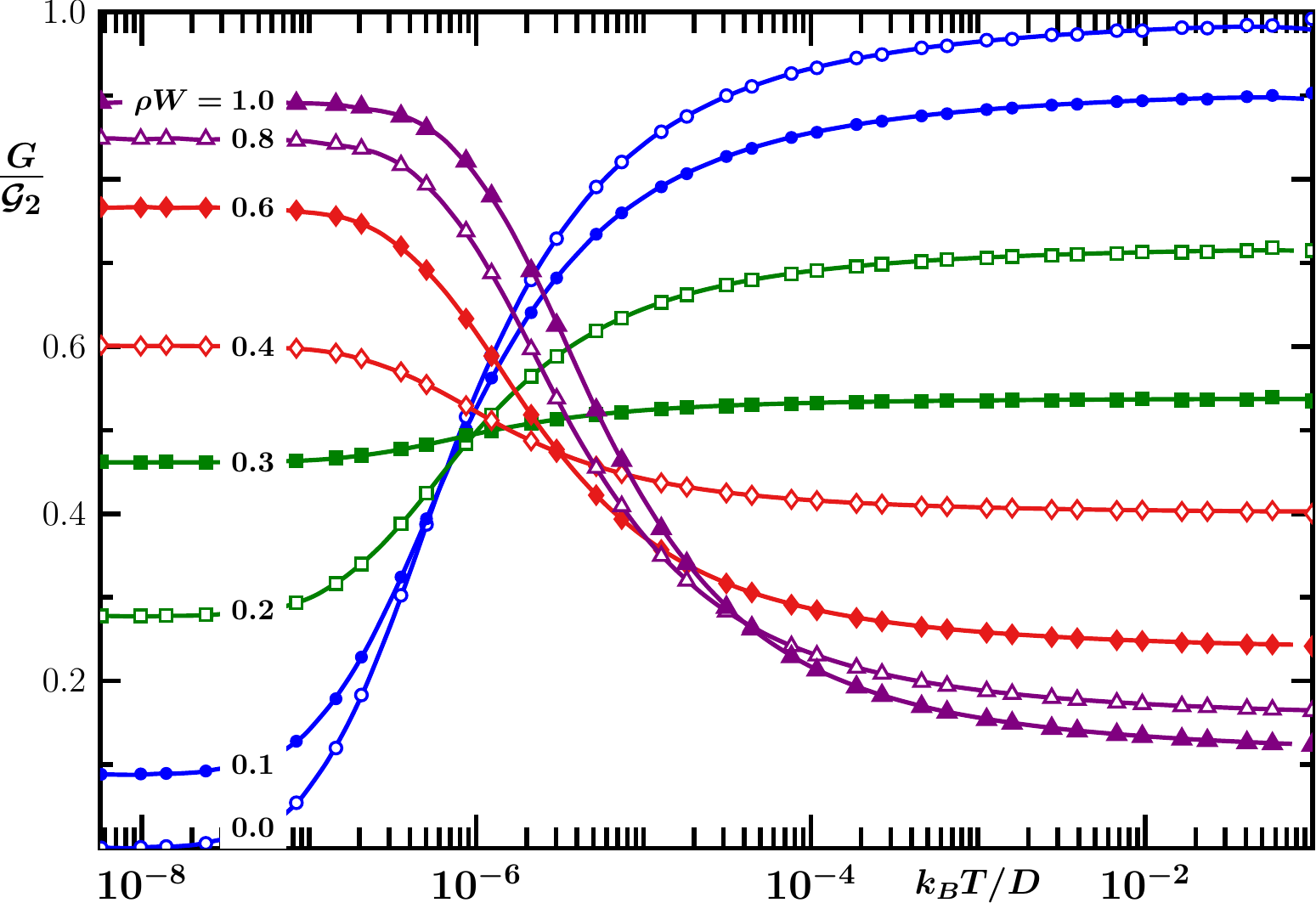}
  \caption{(Color online) Temperature dependence of the conductance
    for $U=5\,D$, $\gammaw=0.15\,D$, $\ed=-2.5\,D$, and the displayed
    wire potentials $W$. The symbols are the results of NRG runs. The
    solid curves through them represent Eq.~(\ref{eq:guniv}) with
    $\delta$ computed from the low-energy eigenvalues of the model
    Hamiltonian; and $T_K$, from the definition $G(T=T_K)=\gc/2$. For
    $W=0$, the model Hamiltonian reduces to the symmetric Hamiltonian,
    and the conductance is complementary to the universal conductance
    curve, $G(T/T_K)=\gc-\gset(T/T_K)$. For larger wire potentials,
    the ground-state phase shifts grow. The agreement with the solid
    lines is nonetheless excellent, because the model Hamiltonians stay
    in the Kondo regime.}
  \label{fig:6}
\end{figure}

Figure~\ref{fig:6} shows the thermal dependence of the conductance for
$\ea=0$ and the eight gate potentials $\rho\vw$ in
Table~\ref{tab:1}. As the potential grows from $\rho\vw=0$ to
$\rho\vw=1$, the ground-state phase shift decreases from $\pi/2$ to
nearly zero, and the conductance $G(T)$ evolves from monotonically
increasing to monotonically decreasing. In all cases, the model
Hamiltonian lying well within the Kondo regime, the agreement with the
solid curves representing Eq.~(\ref{eq:guniv}) is excellent.

\subsubsection{Asymmetric models}
\label{sec:num-thermal-asym}
\begin{figure}[th]
  \centering
  \includegraphics[width=0.95\columnwidth]{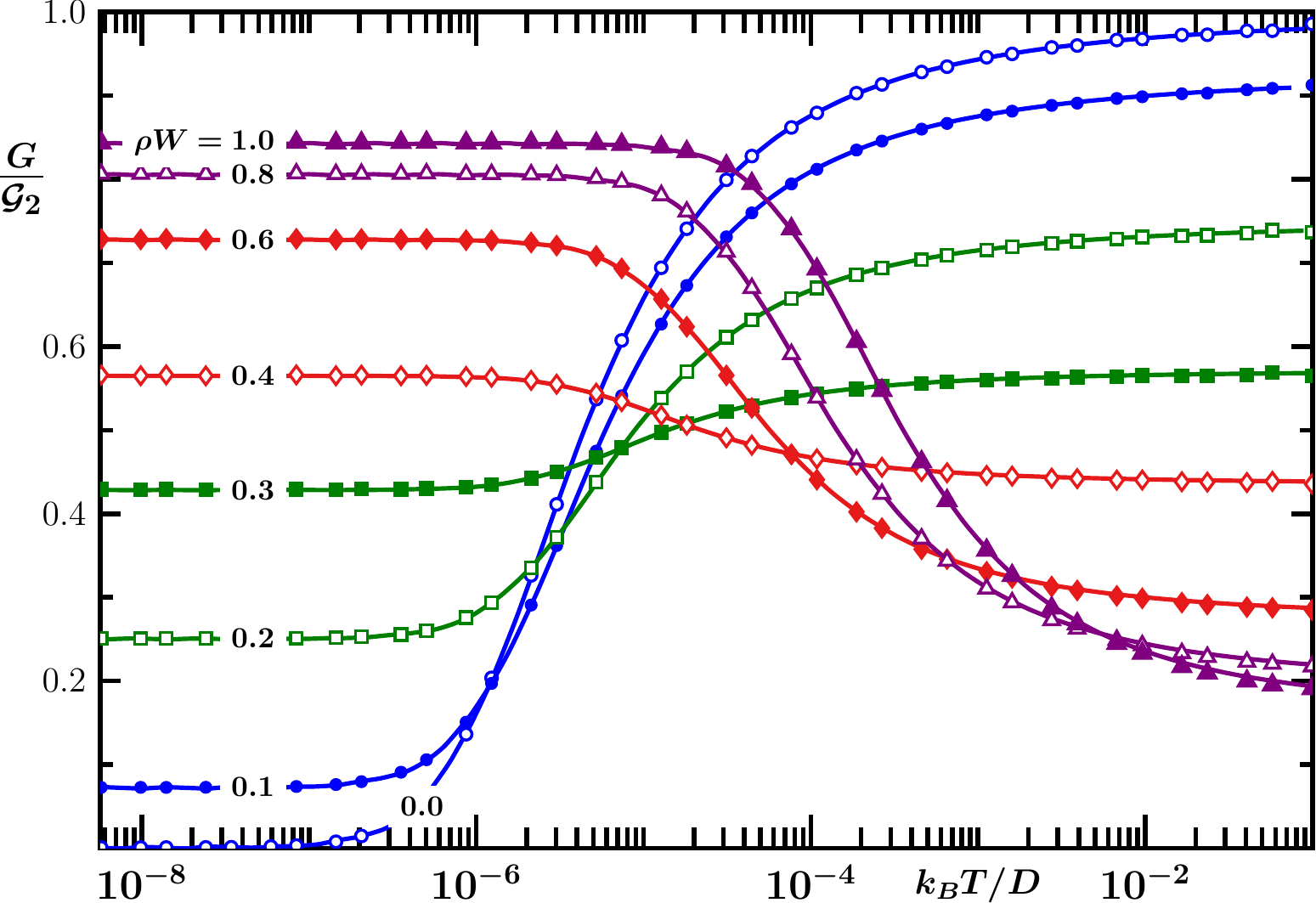}
  \caption{(Color online) Analogous to Fig.~\ref{fig:6}, for
    $\ed=-3.4\,D$. The particle-hole asymmetry, here more pronounced
    than in Fig.~\ref{fig:6}, tends to reduce the conductance at high
    temperatures and enhance it at low temperatures. Since the model
    Hamiltonian for each $\vw$ lies in the Kondo regime, the agreement
    with Eq.~(\ref{eq:guniv}) is again excellent.}
  \label{fig:7}
\end{figure}

Figure~\ref{fig:7} displays temperature-dependent conductances for
$\ed=-3.4\,D$, equivalent to $\ea=-0.9\,D$. The other model parameters
coincide with those in Fig.~\ref{fig:6}. As Table~\ref{tab:1} shows,
the enhanced particle-hole asymmetry makes each ground-state phase
shift somewhat larger than the corresponding $\delta$ in
Fig.~\ref{fig:6}. For $\vw=0$ (open circles) the phase shift in
Fig.~\ref{fig:6} was $\pi/2$, so that the factor $\cos(2\delta)$ on
the right-hand side of Eq.~(\ref{eq:guniv}) is now larger than it
was. The more negative dot energy thus flattens the conductance curve.

For each nonzero potentials, the phase shift associated with
Fig.~\ref{fig:7} in Table~\ref{tab:1} makes the factor $\cos(2\delta)$
smaller than the corresponding factor in Fig.~\ref{fig:6}.
Equation~(\ref{eq:gfl}) therefore pushes down the low-temperature
conductances, while Eq.~(\ref{eq:glm}) pushes up the high-temperature
conductances.  The Kondo temperatures are now spread over a wider
ranger. Apart from such minor changes, however, Figs.~\ref{fig:6}~and
\ref{fig:7} display the same picture. In particular, all Hamiltonians
and temperatures being in the Kondo regime, the numerical conductances
are in very good agreement with Eq.~(\ref{eq:guniv}).

\begin{figure}[th]
  \centering
  \includegraphics[width=0.95\columnwidth]{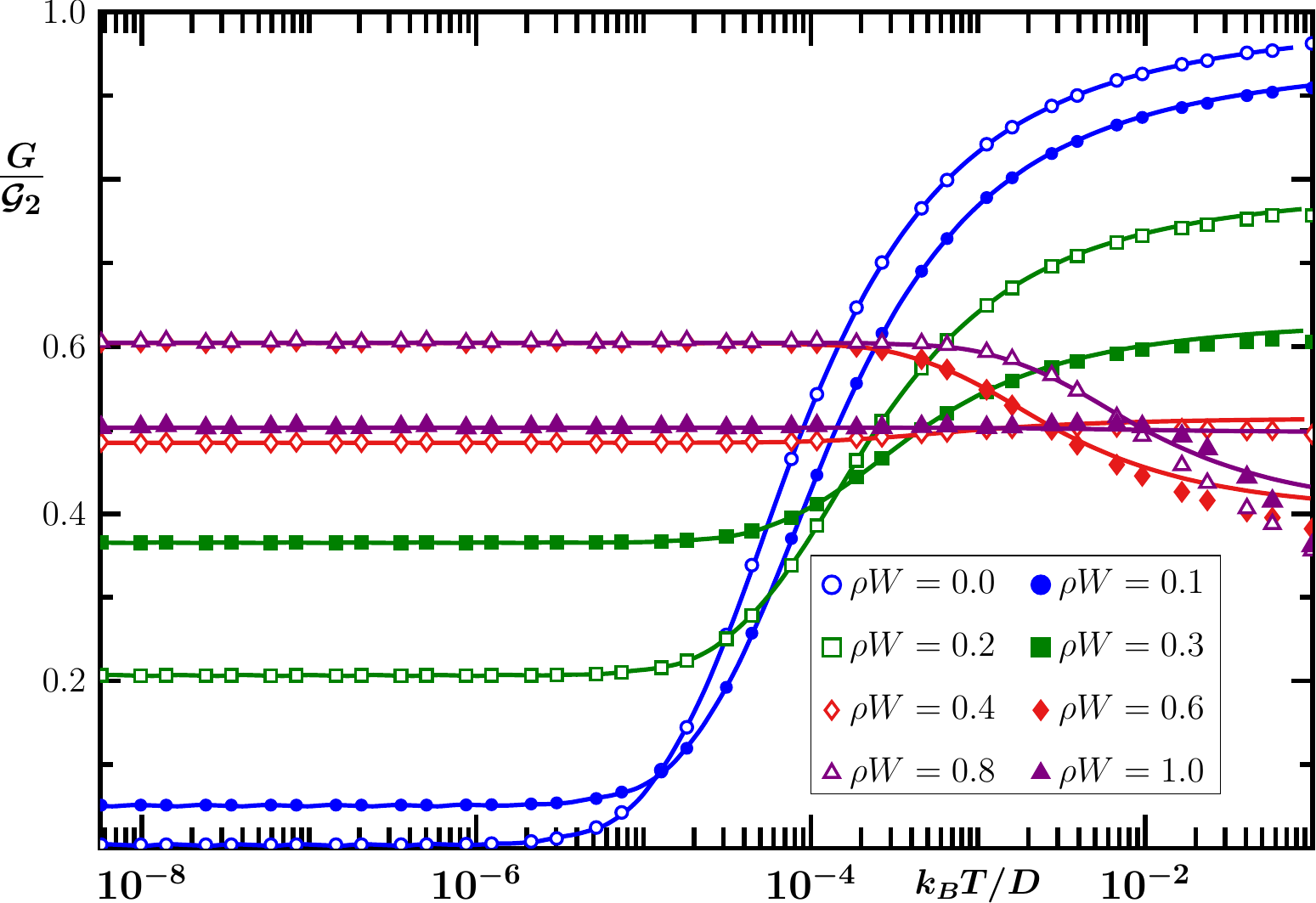}
  \caption{(Color online) Analogous to Figs.~\ref{fig:6}~and
    \ref{fig:7}, for $\ed=-4.0\,D$. Four of the curves, those with
    $\rho W\le0.4$, belong to model Hamiltonians in the mixed-valence
    regime, which restricts Eq.~(\ref{eq:guniv}) to the temperature
    range $k_BT\ll \gammaw$. For $k_BT>10^{-2}\,D$, the diamonds and
    triangles therefore show disagreement with the solid lines. The
    disagreement grows with $W$, because the wire gate potential
    drives the Hamiltonian away from the Kondo regime.}
  \label{fig:8}
\end{figure}

More negative dot energies can displace the Hamiltonian into the
mixed-valence regime. As indicated by the solid lines in
Fig.~\ref{fig:2}, for $\rho \vw=0$ the Kondo domain extends from
$\ea=-(U/2-\Gamma)$ to $\ea=U/2-\Gamma$; with $\ed=-4\,D$, \ie\
$\ea=-1.5\,D$, the model is still within it. As $\rho W$ grows,
however, the Kondo regime is uniformly shifted toward higer dot
energies. It results that the open and filled triangles in
Fig.~\ref{fig:8}, which describe conductances for $\rho W=0.8$ and 1,
respectively, represent Hamiltonians in the mixed-valence
regime. 

While the other curves, computed for smaller $\rho W$'s, agree with
Eq.~(\ref{eq:guniv}), the deviations separating the triangles from the
solid lines at high temperatures, $k_BT\agt10^{-2}\,D$, are
substantial. The discrepancies are reminders that, in the
mixed-valence regime, the dominant characteristic energy is
$\mc{E}_c^{\text{m-v}}=\gammaw$, which restricts the domain of
Eq.~(\ref{eq:guniv}) to $k_BT\ll \gammaw=0.15\,D$. Our numerical study
of the embedded configuration reported similar
deviations.\cite{SYO2009ii} 

Since the conductance curves for large $\rho\vw$ cross the $G=\gc/2$
horizontal at relatively high temperatures, which fall in the
nonuniversal thermal range $k_BT\alt\gammaw$, it would be
inappropriate to extract Kondo temperatures from the identification
$G(T/T_K)=\gc/2$. To evaluate each of the two numbers marked with
asterisks in Table~\ref{tab:1}, we therefore adjusted $T_K$ so that
the corresponding solid line ran through the triangle closest to
$G=0.7\gc$.

\begin{figure}[th]
  \centering
  \includegraphics[width=0.95\columnwidth]{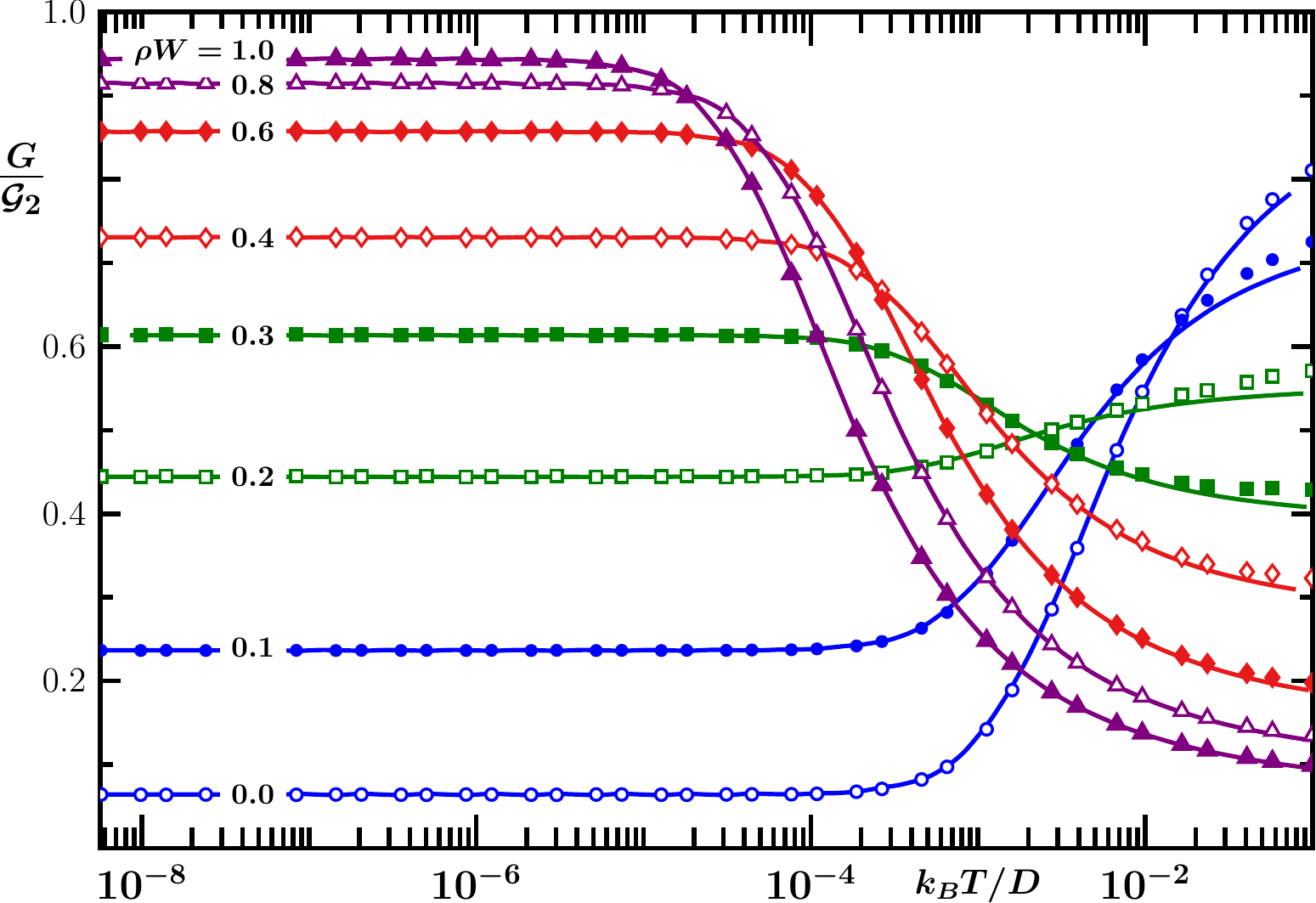}
  \caption{(Color online) Analogous to Figs.~\ref{fig:6}-\ref{fig:8}, for
    $\ed=0.4\,D$.  With $\ed+U/2>0$, the (positive) wire potential
    tends to dampen the effects of particle-hole asymmetry. For $\rho
    W \le 0.3\,D$, the model Hamiltonian now lies in the mixed-valence
    regime, and the calculated conductances deviate significantly from
    the solid lines in the temperature range $k_BT>10^{-2}\,D$.}
  \label{fig:9}
\end{figure}

Since the wire potential displaces the Kondo regime to higher dot
energies, for $\ed\to 0$ we expect the $\rho W=0$ Hamiltonian to leave
the Kondo regime before the $\rho W>0$ Hamiltonians. Accordingly,
Fig.~\ref{fig:9} contrasts triangles and squares very well fitted by
Eq.~(\ref{eq:guniv}) with circles and diamonds that depart
significantly from the solid lines representing the universal mapping
for $k_BT>3\times10^{-2}\,D$.

The ground-state phase shift $\delta$ decreases with $\ed$. Thus,
while the phase shift for each curve in Figs.~\ref{fig:7}~and
\ref{fig:8} exceeds the phase shift for the same $\rho W$ in
Fig.~\ref{fig:6}, the phase shift for each curve in Fig.~\ref{fig:9}
is smaller than the corresponding phase shift in Fig.~\ref{fig:6}. In
compliance with Eq.~(\ref{eq:gfl}), the low temperature conductances
in Fig.~\ref{fig:9} are higher than the corresponding $G(T\to0)$ in
Fig.~\ref{fig:8}. At high temperatures, the analogous comparison is
unproductive, because the relatively high Kondo temperatures place the
high-temperature end of the axis in Fig.~\ref{fig:9} outside the reach of
Eq.~(\ref{eq:glm}).

\subsection{Comparison with experiment}
\label{sec:num-exp}
The numerical results in Figs.~\ref{fig:3}-\ref{fig:5} can be compared
to a number of experimental studies of transport in $T$-shaped
devices.\cite{KAS+04:035319,SAK+05:066801,KSA+06:36,%
  AEO+06:195329,OAK+07:084706,Kat07:233201} Particularly illustrative
are Fig.~1(c) in Ref.~\onlinecite{SAK+05:066801}, and Fig.~5 in
Ref.~\onlinecite{KSA+06:36}. The former, which shows low-temperature
conductance profiles for various level widths, exhibits valleys,
plateaus, and antiresonances---all the conspicuous features in the
conductance profiles at $k_BT=10^{-10}\,D$ in Figs.~\ref{fig:3}-\ref{fig:5}.

Even more instructive is the illustration in
Ref.~\onlinecite{KSA+06:36}, which demonstrates that the wire gate
potential controls the division of the current between the wire and
the bypass. The evolution of the low-temperature conductance profile
under an increasing wire gate potential is documented in the plot. The
profile starts out as Kondo valley analogous to the low-temperature
profile in Fig.~\ref{fig:3}. It then evolves to a curve marked by
interference, analogous to the low-temperature profile in
Fig.~\ref{fig:4}(c). Finally, it reaches a Kondo plateau, analogous to
the low-temperature profile in Fig.~\ref{fig:5}(c).  The NRG results
and Eq.~(\ref{eq:guniv}) are therefore in qualitative agreement with
the experimental results.

What is more important, the mapping describes quantitatively the
thermal dependence of measured conductances. A fit of the 
temperature-dependent conductances resulting from two gate potentials
applied by Sato et al.\cite{SAK+05:066801} to a $T$-shaped device was
presented in Ref.~\onlinecite{SYO09:000}. A 
background current being detected, three parameters were involved in
the fit to the first conductance curve: the background conductance,
the phase shift, and the Kondo temperature $T_K$. The same first two
parameters and a different Kondo temperature then fitted the second
curve. In each case, within the dispersion of the experimental data,
optimum agreement resulted.

\section{Conclusions}
\label{sec:concl}

Equation~(\ref{eq:guniv}) offers a unifying view of electrical
conduction through a quantum wire side-coupled to a quantum dot. Valid
over the entire Kondo regime, it captures with error $\mc{O}(k_BT/D)$
the conductance crossover from the local-moment regime (\ie\
high-temperatures, $T\gg T_K$), equivalent to a conduction band weakly
interacting with the dot magnetic moment, to the low-temperature
regime ($T\ll T_K$), in which the dot electron and the conduction
electrons around it lock into a singlet that reduces the model
Hamiltonian to a phase-shifted conduction band.

The phase shift controls the thermal dependence of the
conductance. With no wire gate potential, the Friedel makes
$\delta\approx\pi/2$, and the Kondo cloud lowers the conductance from
nearly ballistic at high $T$ to nearly zero at low $T$. For $\vw=0$, the
conductances for the $T$-shaped device and the single-electron
transistor are complementary.\cite{SYO2009ii} 

A (positive) gate potential applied to the wire in the $T$-shaped
device induces electric charge that reduces the ground-state phase
shift. As $\rho W$ pushes $\delta$ below $\pi/4$, the conductance
curve first becomes flat and then reverses the pattern in
Fig.~\ref{fig:3}.  For large $\rho\vw$, as indicated by
Fig.~\ref{fig:5}(c), the conductance through the $T$-shaped device
approaches the conductance through a single-electron
transistor.\cite{SYO2009ii}

The Friedel sum rule explains why the two devices respond differently
to the wire gate potential. In the embedded configuration, since the dot
is in the conduction path, its occupation $n_d$ controls the
conductance. Although the wire gate potential $\vw$ determines the
amount of charge in the wire, it has little influence upon $n_d$:
according to the Friedel sum rule, $n_d=2(\delta-\deltaw)/\pi$, so
that the charge $2\deltaw/\pi$ induced by the gate potential $\vw$ is
subtracted from the total charge $2\delta/\pi$ piled up in the quantum
wire. We therefore expect the conductance $G$ through the
single-electron transistor to be independent of $\vw$. The mapping
between $G(T/T_K)$ and the universal function $\gset(T/T_K)$, which is
parametrized by the phase-shift difference $\delta-\deltaw$, ractifies
this reasoning.\cite{SYO2009,SYO2009ii}

By contrast, the mapping between the conductance through the device in
Fig.~\ref{fig:1} and the universal function $\gset(T/T_K)$ is
parametrized by the phase shift $\delta$. According to the Friedel sum
rule, the wire charge is $2\delta/\pi$. Much as the dot charge
$n_d=2(\delta-\deltaw)/\pi$, which controls conduction in the embedded
configuration, parametrizes the mapping to the conductance through the
single-electron transistor,\cite{SYO2009} the wire charge
$2\delta/\pi$ controls conduction in the side-coupled configuration
and parametrizes the mapping~(\ref{eq:guniv}). While the dependence on
the dot charge practically shields the single-electron transistor from
gate potentials applied to the wire, the dependence on the wire charge
makes the conductance through the $T$-shaped device remarkably sensitive
to such potentials.

The wire gate potential affects Fig.~\ref{fig:1} in two ways. First,
the total charge in the $\phi_0$ orbital, equal to the charge induced
by the wire potential plus, at low temperatures, the charge in the
Kondo droplet, obstructs transport through the central portion of the
wire. Second, the potential couples $\cd$ to states orthogonal to
$\phi_0$ and thus opens a bypass through the quantum dot. For very
weak (strong) potentials, the path through $\phi_0$ ($\cd$) is
dominant; the landscape in Fig.~\ref{fig:3} [~\ref{fig:5}(c)] is thus
complementary (very similar) to the landscape for a single-electron
transistor.\cite{SYO2009ii}

For intermediate potentials [Fig.~\ref{fig:4}(c)], the interference
between the currents along the two paths makes the landscape markedly
asymmetric in the mixed-valence region. At the same time, in the Kondo
regime, the two parallel conduction paths, one efficient at low
temperatures, the other efficient at high temperatures, make the
landscape remarkably flat.

The practical value of universality has been demonstrated in both
experimental arrangements. In the embedded configuration, early in the
history of the single-electron transistor, the universal function
$\gset(T/T_K)$ guided the interpretation of conductance
data.\cite{GGK+98:5225} In the side-coupled arrangement,
Eq.~(\ref{eq:guniv}) was shown to fit conductance curves generated in
the laboratory\cite{SAK+05:066801,KSA+06:36} accurately enough to
determine the Kondo temperature and ground-state phase shift within
deviations set by the dispersion of the experimental data.\cite{SYO09:000}

Neither the universal mapping~(\ref{eq:guniv}), nor the corresponding
expression for the embedded configuration\cite{SYO2009} determine
explicitly any of the model parameters, let alone the physical
constants emulated by the model Hamiltonian. They offer indirect
information that may assist future {\em ab initio} descriptions of the
physical properties of single-electron transistors or $T$-shaped
devices. The two mappings of the Kondo-regime conductances to the
universal function $\gset(T/T_K)$ redefine the ultimate target of such
{\em ab initio} problems, from the computation of conductance curves
to the computation of phase shifts and Kondo temperatures.  In a less
challenging arena, the mappings bring Kondo-regime conductance curves
within the reach of the Bethe-{\em ansatz}
approach.\cite{AFL83:331,TW83:453} Finally, on the basis of the
experience with thermodynamical properties,\cite{SLO+96:275,CO04.01}
we expect the mappings to offer benchmarks to check the accuracy of
numerically or analytically computed transport properties for model
Hamiltonians describing quantum-dot arrays.

\acknowledgments One of us (LNO) thanks the hospitality at the
Institut Henri Poincaré (Paris), where this research was started. This
work was supported by the CNPq and FAPESP.


\end{document}